# Mass, Speed, Direction:
# John Buridan's 14th century concept of momentum

Christopher M. Graney
Jefferson Community & Technical College
1000 Community College Drive
Louisville, KY 40272
christopher.graney@kctcs.edu

Abstract:

In the 14th century the French thinker John Buridan developed a theory of motion that bears a strong resemblance to Newtonian momentum. Buridan's ideas include a quantity of motion which is determined by an object's mass, speed, and direction; in the absence of resistive effects, this quantity remains with the object. Buridan's work is an interesting story in the history of physics. Buridan's insights have value for introducing concepts of inertia and momentum to physics students.

"**M**odern science began in the Middle Ages," a fact that has been forgotten thanks to the celebrated accomplishments of Copernicus and Galileo, who did not acknowledge their predecessors. So states James Hannam in a January 2010 article in *History Today*. Among the scientists of the Middle Ages that Hannam mentions is John Buridan, a French thinker who was the first to develop modern concepts of inertia and momentum.[1] Buridan's work has been known to historians of science for decades,[2] and remains a topic of discussion among them today.[3,4] However, it is not well-known in physics circles,[5] although there was an *American Journal of Physics* discussion of Buridan 35 years ago as part of a history of inertia.[6] Readers of *The Physics Teacher* may find Buridan of interest both as a matter of history and because Buridan presents important physics ideas in a different sort of way which may be of value in the physics classroom.

John Buridan lived in the first half of the 14th century, and was for a time the rector of the University of Paris.[7] His work in physics focused on the question of what keeps an object such as a tossed stone or a freely spinning wheel, moving once set in motion. Aristotle, whose ideas dominated physics until the time of the scientific revolution, said motion requires a mover: when a person drags a chair across the room, the person is the chair's mover; when the chair is released, it ceases motion. When a person throws a rock, the situation is more complex: the person is the mover until the rock is released, at which point the air through which the rock moves takes over as mover, and carries the rock along.[8] Buridan argued that a projectile is not moved by the air, because the air resists, not causes, a projectile's motion.[9] He used both physical experiments and thought experiments to show this to be true. He proposed his own theory of motion, which looks very much like something a physicist would teach in class today. And finally, he used his theory to explain specific types of motion — explanations which again look very much like something from a physics class today.

Buridan quotes Aristotle as saying that —

> Projectiles are moved further after the projectors are no longer in contact with them, either by antiperistasis [the air rushing in behind the moving projectile to replace the void it creates by its motion], as some say, or by the fact that the air having been pushed, pushes with a movement swifter than the movement of impulsion by which it (the body) is carried towards its own place [its "natural" place – for Aristotle, bodies like rocks naturally moved towards the center of the Earth, just like they were naturally dense].[10]

Then Buridan says that experiments with a spinning top or with a "smith's mill" (a grinding wheel – Figure 1) show that this cannot be true. A top or a grinding wheel, once set moving, continues to move for long time. Since it spins in place it does not displace air; it leaves no void as it moves. So, Buridan argues, it makes no sense to say that a top or wheel is kept moving by the action of air.  Also, a ship rowed swiftly through still water continues moving even after the rowers cease, he says, and yet people in the ship feel no air from behind them pushing the ship along; they feel the resistance of air from the front.[11]

Next, Buridan describes some simple experiments to test whether air is the source of continued motion of a grinding wheel or a ship:

> [I]f you cut off the air on all sides near the smith's mill by a cloth, the mill does not on this account stop but continues to move for a long time.  Therefore, it is not moved by air.

> [I]f the ship were loaded with grain or straw and were moved by the ambient air, then that air ought to blow exterior stalks toward the front.  But the contrary is evident, for the stalks are blown rather to the rear because of the ambient air.

Finally, Buridan extends his criticism of Aristotle directly to projectiles.  If air is what moves a projectile —

> [I]t follows that you would throw a feather farther than a stone and something less heavy farther than something heavier, assuming equal magnitudes and shapes.  Experience shows this to be false.

Having laid out both his general criticism of, and some experiments for illustrating the problems with, Aristotle's ideas, Buridan offers his own theory of motion:

> Thus we can and ought to say that in the stone or other projectile there is impressed something which is the motive force of that projectile.  And this is evidently better than falling back on the statement that the air continues to move that projectile.  For the air appears rather to resist.  Therefore, it seems to me that it ought to be said that the motor [not a motor in the modern sense, but the source of motion, such as the hand that launches the stone] in moving a moving body impresses in it a certain momentum or a certain motive force of the moving body, in the direction toward which the mover was moving the moving body, either up or down, or laterally, or circularly.  And by the amount the motor moves that moving body more swiftly, by the same amount it will impress in it a stronger momentum.  It is by that momentum that the stone is moved after the projector ceases to move.[12]

To this Buridan adds that

> …by the amount more there is of matter, by that amount can the body receive more of that momentum…

Buridan's concept appears quite similar to the modern concept of momentum: the quantity of momentum given to a body is (1) proportional to its speed, (2) in the direction of motion, (3) proprotional to the amount of matter in the body.  The primary difference between this and the modern concept of momentum seems to be that Buridan does not specifically distinquish between linear and rotational momentum.

Buridan describes air as being resistive. He combines this idea with his momentum idea to explain what we would call "projectile motion with air resistance":

> It is by that momentum that [a thrown stone] is moved after the projector ceases to move. But that momentum is continually decreased by the resisting air and by the gravity of the stone, which inclines it in a direction contrary to that in which the momentum was naturally predisposed to move it. Thus the movement of the stone continually becomes slower, and finally that momentum is so diminished or corrupted that the gravity of the stone wins out over it and moves the stone down to its natural place [Buridan continues to accept the idea that heavy objects fall to Earth due to their natural properties – their "gravity"].

Buridan seems to be saying that the motion of a thrown stone depends on its initial momentum, the effect of air resistance which acts to reduce that momentum, and the effect of gravity which changes its direction. A modern verbal description of projectile motion with air resistance would say that the stone is given an initial momentum when launched; during its flight air resistance will continually act to decrease the magnitude of the stone's momentum, while gravity will continually act to increase the downward component of the stone's momentum. The net effect will be to reduce the stone's momentum and to deflect the stone's momentum vector downward.

The reader may feel that Buridan's description makes it sound as though projectile motion is a two-step process — in the throw the stone is given momentum; that momentum is lost, and then gravity takes over and brings the stone to Earth — and such as two-step process is not modern momentum. But, consider Buridan's discussion of the motion of more dense and less dense projectiles:

> A feather receives [momentum] so weakly that such momentum is immediately destroyed by the resisting air. And so also if light wood and heavy iron of the same volume and of the same shape are moved equally fast by a projector, the iron will be moved farther because there is impressed in it a more intense momentum, which is not so quickly corrupted as the lesser momentum would be corrupted. This is also the reason why it is more difficult to bring to rest a large smith's mill which is moving swiftly than a small one, evidently because in the large one, other things being equal, there is more momentum. And for this reason you could throw a stone of one-half or one pound weight farther than you could a thousandth part of it. For the momentum in that thousandth part is so small that it is overcome immediately by the resisting air.

Here Buridan is spot on in his discussion of projectile motion with air resistance. Two projectiles of equal size and shape but differing densities launched at identical speeds will behave exactly as Buridan describes, for the reasons he describes (Figures 2 and 3). Usually modern physicists do not describe such motion in terms of momentum, but we certainly can do so.

Buridan also applies his momentum theory to dense objects dropped to the ground from rest:

> From this theory also appears the cause of why the natural motion of a heavy body downward is continually accelerated. For from the beginning only the gravity was moving it. Therefore, it moved more slowly, but in moving it impressed in the heavy body a momentum. This momentum now together with its gravity moves it. Therefore, the motion becomes faster; and by the amount it becomes faster, so the momentum becomes more intense. Therefore, the movement continually becomes faster.

Again, physicists usually do not describe free fall in terms of momentum, but we can. A modern discussion of free fall couched in terms of momentum is very similar to Buridan's discussion of free fall in which gravity continually increases the momentum of the falling body.

Buridan even applies his theory to a case where there would be no resistance at all — the case of celestial bodies. Buridan says that it could be argued that —

> …God, when He created the world, moved each of the celestial orbs as He pleased, and in moving them He impressed in them momenta which moved them without his having to move them any more… [and the momenta of the] celestial bodies were not decreased nor corrupted afterwards, because there was no inclination of the celestial bodies for other movements. Nor was there resistance which would be corruptive or repressive of that momentum.

In short, in the one case where he can envision no corruption of momentum, Buridan says that an object in motion remains in motion indefinitely.

Lastly, Buridan even applies his theory to more complex cases, such as bouncing balls and vibrating strings. A ball he describes as being elastic like an archer's bow, and so when it is thrown hard to the ground it is compressed under its own momentum, and its elastic return to a spherical shape propels itself back upward, giving itself an upward momentum that carries it up a distance. He says a string bent in one direction will return swiftly to its normal straight position, "But on account of the momentum, it crosses beyond the normal straight position in the contrary direction, and then returns. It does this many times." He says the pendulum motion of a swinging bell works the same way, and notes that a heavy swinging bell cannot easily be stopped. His discussions of these cases are consistent with a Newtonian treatment of them, if done from the perspective of momentum. Historians of science have recognized Buridan's work on oscillating bodies as being "as profound as Galileo's discovery of the law of isochrony two and a half centuries later."[13]

Whether Buridan's ideas were used by Galileo and not acknowledged as Hannam states is a matter for historians of science to determine. The well-known historian of science Pierre Duhem argued that Buridan's exact words echo in the writings of 17th-century scientists working on mechanics,[14] and the author of the previously mentioned *AJP* article that discusses Buridan agrees that the work of Buridan influenced scientists such as Galileo.[15] But to a physicist, it is fascinating just to see these ideas in play centuries before what most of us think of as the dawn of modern science.

Buridan's story is a fun tale to tell to students. Moreover, Buridan's discussions are so insightful that they suggest innovative ways of presenting the concept of momentum to students who may resist the idea that "an object in motion remains in motion". Buridan was writing in a time before modern algebra had come into use. If his descriptions seem vivid, perhaps that is because in his day verbal description of ideas in physics was more common. Moreover, he is expressing ideas without the benefit of a training in Newtonian physics, something he shares in common with the introductory physics student!

James Hannam is right that we should acknowledge that modern science began with thinkers like Buridan, whose ideas predated by centuries (and perhaps influenced) the ideas of the more familiar scientists such as Galileo and Copernicus. And Buridan's ideas provide interesting insight into physics which can be of use in reaching introductory physics students. Next semester, try building a low-friction wheel and introduce your students to some ideas about physics through John Buridan!

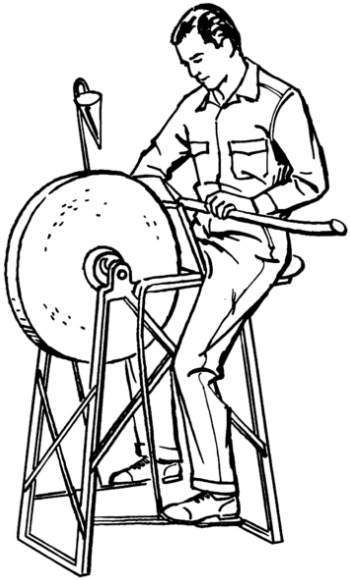

**FIGURE 1**

**A smith's mill or grinding wheel (from Pearson Scott Foresman, Wikimedia Commons).**

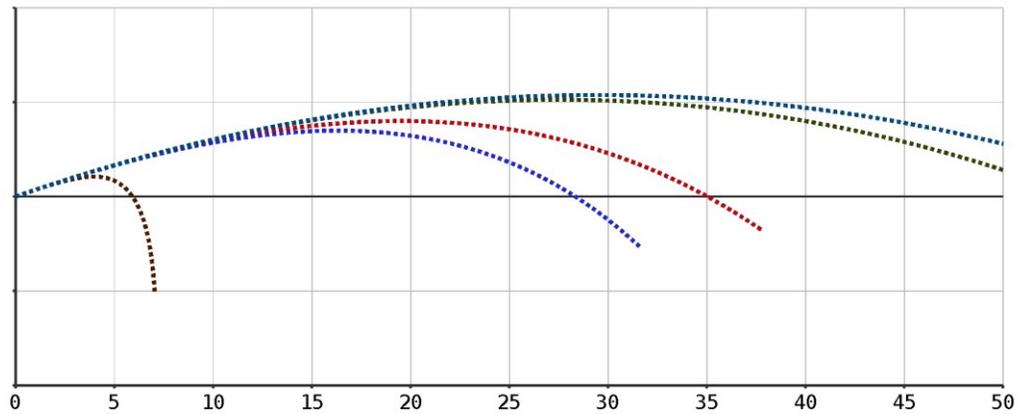

**FIGURE 2**

Trajectories for projectiles (all rough spheres with diameter 5 cm) launched at a velocity of 30 m/s at 20°above the horizonal. Units are meters, with same scale on both axes. Projectiles differ in specific gravity (s.g.). From topmost curve to bottommost – no air resistance, iron projectile (s.g. = 8), watery projectile (such as an apple, s.g. = 1), wood projectile (s.g. = 0.6), hollow ball (s.g. = 0.05). The trajectories show 2 seconds of the projectile's motion (the top two are truncated for reasons of space). Less dense projectiles travel much less far than more dense ones.

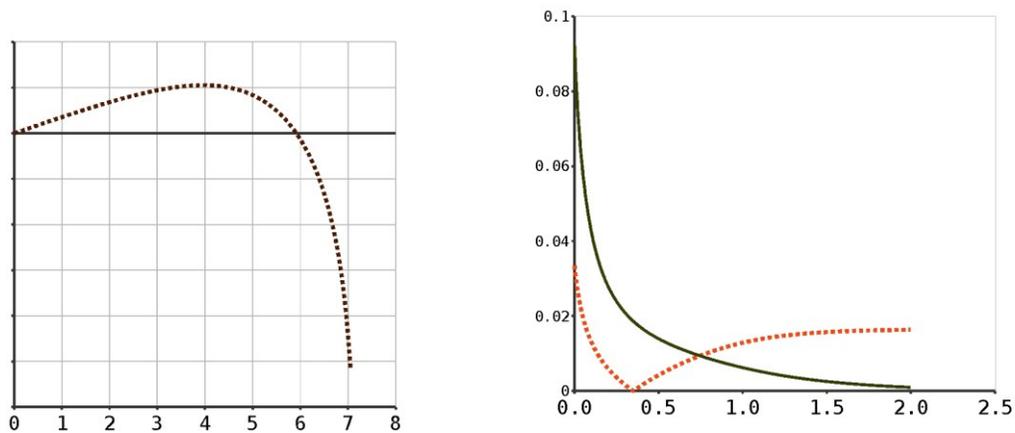

**Figure 3**

**Left – trajectory for the least dense projectile in Figure 2. Units are meters, with same scale on both axes. Right – plot of magnitudes of momentum (in kgm/s) vs. time (s) for horizontal (solid) and vertical (dotted) components.**

For a projectile in which air resistance is very significant, the projectile briefly travels along a relatively flat trajectory, but as air resistance robs it of its initial momentum, it quickly transitions to vertical motion. Buridan's statement that, after a projectile is launched its movement continually becomes slower until its momentum is so diminished that gravity wins out over it and moves it downward, is a reasonable description of the motion of a projectile with significant air resistance. Projectiles such as feathers have their initial momentum immediately dissipated by air resistance and flutter downward under the influence of gravity in a manner much like Buridan's description.

[1] James Hannam, "Lost pioneers of science," *History Today* **60**, 5-6 (January 2010). See also <http://www.historytoday.com/james-hannam/lost-pioneers-science>.

[2] Marshall Clagett, "Some general aspects of physics in the middle ages," *Isis* **39**, 29-44 (1948).

[3] Abel B. Franco, "Avempace, projectile motion, and impetus theory," *Journal of the History of Ideas* **64**, 521-546 (2003).

[4] J.M.M.H. Thijssen, "The Buridan School Reassessed: John Buridan and Albert of Saxony," *Vivarium* **42**, 18-43 (2004).

[5] For example, a search for Buridan on "Scitation" <http://scitation.aip.org/> on 2 April 2013 yielded no relevant returns at all.

[6] Allan Franklin, "Principle of Inertia in the Middle Ages," *American Journal of Physics* **44**, 529-545 (1976).

[7] "John Buridan" in *A Source Book in Medieval Science*, edited by Edward Grant (Harvard University Press, MA, 1974), p. 819.

[8] A thorough discussion of Aristotle's ideas is found in Franklin, "Principle of inertia…"

[9] John Buridan, "The impetus theory of projectile motion, from 'Questions on the Eight Books of the Physics of Aristotle'" (translated by M. Clagett), in *A Source Book in Medieval Science*, edited by Edward Grant (Harvard University Press, 1974), p. 275.

[10] Buridan, "The impetus theory…" p. 275. For a slightly different translation of Aristotle's statement on projectiles, see Franklin, "Principle of inertia…" p. 530.

[11] This and all following Buridan quotes and paraphrases are from Buridan, "The impetus theory…" p. 275-278 (available via Google Books at <http://books.google.com/books?id=fAPN_3w4hAUC>).

[12] Buridan's original work was in Latin, and he used the Latin word "impetus" as shorthand for the "motive force of a moving body". Clagett's translation leaves the term "impetus" in place, but in this paper I replace it with "momentum," as that is the physics concept that Buridan's concept resembles. Franklin ("Principles of inertia…") says that to equate Buridan's idea to Newtonian momentum would be "a gross anachronism [p. 538]." I leave it to reader to decide if that is true from an historian's perspective. From a physics perspective, however, what are of interest are the theory that is being described, how well the theory works, and its potential value in the class room.

[13] Bert S. Hall, "The scholastic pendulum," *Annals of Science* **35**, 441-462 (1978), pp. 447-448.

[14] Pierre Duhem, "Research on the history of physical theories," *Synthese* **83**, 189-200 (1990, originally published in French in 1917) p. 192.

[15] Franklin, "Principle of inertia…" p. 540, 541.